# ON THE MODELING OF OPENFLOW-BASED SDNS: THE SINGLE NODE CASE

Kashif Mahmood[1], Ameen Chilwan[2], Olav N. Østerbø[1] and Michael Jarschel[3]

[1]Telenor Research, Norway
{kashif.mahmood,olav-norvald.osterbo}@telenor.com
[2]Department of Telematics, NTNU, Norway
chilwan@item.ntnu.no
[3]Nokia, Germany
michael.jarschel@nsn.com

## ABSTRACT

*OpenFlow is one of the most commonly used protocols for communication between the controller and the forwarding element in a software defined network (SDN). A model based on M/M/1 queues is proposed in [1] to capture the communication between the forwarding element and the controller. Albeit the model provides useful insight, it is accurate only for the case when the probability of expecting a new flow is small.*

*Secondly, it is not straight forward to extend the model in [1] to more than one forwarding element in the data plane. In this work we propose a model which addresses both these challenges. The model is based on Jackson assumption but with corrections tailored to the OpenFlow based SDN network. Performance analysis using the proposed model indicates that the model is accurate even for the case when the probability of new flow is quite large. Further we show by a toy example that the model can be extended to more than one node in the data plane.*

## KEYWORDS

*OpenFlow, Performance analysis, Queuing theory, Software defined networks.*

## 1. INTRODUCTION

Software defined networking (SDN), an academic lead initiative, has already made a lot of impact in the datacenters. As early as January 2012, Google had their full scaled datacenter WAN running as OpenFlow based SDN [2]. SDN is now all set to roar in the carrier networks domain too. This is because SDN promises network deployment and service upgrade on software which has huge benefits for the network operators because in the future the network operators will not compete on the basis of network coverage alone but on the basis of features and services.

All this has been possible due to the basic architectural principle of SDN which is the separation of the control plane from the data plane. The architecture involves SDN controller(s) residing in the control plane while the forwarding element(s) make the data plane. In order to handle the communication between the control plane and the data plane elements, OpenFlow is the only **open, standard** protocol [3].

OpenFlow started as a test protocol in Stanford but is now managed and maintained by Open Networking Foundation (ONF) [3]. It started with OpenFlow version 1.0.0 and at the writing of the paper, version 1.4 has been specified [3]. The working principle is the same but each version involves some additional features. For example the version 1.1.0 has support for group tables which was not there in version 1.0.0. The work in this paper is based on OpenFlow version 1.0.0 and we believe that it can be easily extended to the new versions.

Under an OpenFlow network, the controller-to-switch communication takes place as follows: where we use the term switch and node interchangeably to represent the forwarding element in the data plane in an SDN network.

When a *flow* with no specified forwarding instructions comes into a network the following actions are taken:

i. A packet (or part of the packet) of the flow is sent by the switch to the controller, assuming that the switch is not configured to drop unknown packets.

ii. The controller computes the forwarding path and updates the required nodes in the data path by sending entries to be added to the flow tables.

iii. All subsequent packets of the flow are forwarded based on pre-calculated forwarding decisions and do not need any control plane action.

It is important to model the controller-to-switch communication for the performance analysis of OpenFlow (OF)-based SDN networks. The modeling of OpenFlow networks will help us to answer questions such as how much data we can pump into the network, what is the packet sojourn time, when and what (switch or the controller) is the bottleneck in a network.

Most of the work on performance analysis of SDN networks is based on simulations or experimentations. Albeit their benefits, analytical modeling is a time efficient alternative because setting up an SDN experiment or performing a simulation can take hours. But the real strength of analytical model lies in the extent to which it can be used for analysing networks and confidence that could be put in the obtained results.

The analytical model should be able to capture actual OpenFlow working principle and at the same time shall be flexible to handle any amount of query traffic going to the controller. Further, the model shall be readily extendable to more than one node in the data plane.

The analytical modeling of OpenFlow-based networks has only been attempted in a handful of papers before. For example feedback oriented queuing theory has been used in [1] to capture the control plane and data plane interaction where the Markovian servers are assumed for both the controller and the switch. However the model becomes less accurate as the probability of traffic going to the controller increases. Secondly it is not clear how the model can be extended to more than one switch in the data plane.

In [4], a network calculus based approach is used to quantify the packet processing capability of the switch in the data plane. However the feedback between the nodes in the data plane and the controller is not considered. This shortcoming of feedback modeling is addressed in [5]. However the model is depicted only for a single node in the data plane and the time stopping method employed therein has limited real time application. Secondly the framework used in [4] and [5] is based on ***deterministic*** network calculus which does not provide any meaningful bounds [6]. To the best of our knowledge apart from these handful analytical works, almost all the other efforts of evaluating performance of OpenFlow-based networks are carried out by simulations or measurements, for example [7], [8], [9]. Moreover, Cbench tool to benchmark the controller performance is also introduced in [10] and is proved to be instrumental in benchmarking.

It is therefore of paramount importance to have an analytical model which can capture the feedback interaction between the controller and the switch, is able to model any amount of traffic going from switch to controller (and vice versa), and can be easily extended to more than one switch in the data plane. The model proposed in this paper is an attempt in that direction.

We model the OpenFlow network as a Jackson network but with a modification to accurately represent the traffic flow from the switch to the controller in an actual OF-based SDN network.

It is highlighted later in the paper (Fig. 2) that this modification to the native Jackson network is necessary to capture the OpenFlow working principle. The model is in turn used for performance analysis of OF-based SDN networks to calculate the mean packet sojourn time and to find out how much data we can pump into the network. The main contributions of this work are:

- A model is proposed to capture the feedback interaction between the switch and the controller mimicking an actual OpenFlow based SDN network

- The model is accurate even for the case when large amount of new flows are arriving at the switch.

- The model can be easily extended to more than one switch in the data plane.

- We show mathematically that the packet sojourn time calculated by our proposed model based on Jackson assumption is the same as the one explicitly calculated for OF-based SDN network in [1].

The rest of the paper is organized as follows; we first present the system model along with the limitations and the necessary preliminaries in Section 2. The performance measures are outlined in Section 3, while the numerical results are presented in Section 4. An insight as to how the proposed model can be used for multi-node case is highlighted in Section 5, while the conclusions along with the future research directions are presented in Section 6.

## 2. MODEL DESCRIPTION

We assume that the overall traffic process at the switch and at the controller follows Poison process similar to [1] given that the two processes are on a different time scale. Further we assume Markovian servers for the switch and the controller wherein we incorporate the transmission time of the packets from the switch to the controller in the service time of the controller. As for the buffer size we assume infinite buffer for the switch and the controller.

We use Jackson network model to represent the OF-based SDN network. To this end a recap of the Jackson model for open queuing networks [11], in which the nodes behaves locally as single M/M/1 queues, is outlined. Albeit trivial, this is done in order to highlight that Jackson model *cannot* be used as such for modeling OF-based SDNs.

Let us consider a Jackson network consisting of two nodes $1$ and $c$ connected in a feedback path as shown in Fig. 1(a). The service rates of nodes $1$ and $c$ are exponentially distributed with average values of $\mu_1$ and $\mu_c$, respectively. The external traffic arrival to the node $1$ is denoted as $\lambda_1$ packets per unit time.

Let $\Gamma_1$ be the net input to node $1$ out of which $\Gamma_c = q_1^{jack} \Gamma_1$ goes to the node $c$ where $q_1^{jack}$ is the probability that the packet goes to the node c.

Assuming that no packet is lost at the controller (infinite buffer at the controller) the balance equation for the system can be written as

$$\Gamma_1 = \lambda_1 + q_1^{jack} \Gamma_1 \qquad (1)$$

It is the term $q_1^{jack}$ which needs modification in order to model OF-based SDN in Fig. 1(b) as a Jackson network in Fig. 1(a). This is because an OF-based SDN, as shown in Fig. 1(b), has the following two salient features:

i. A packet coming to any node in the data plane will at most visit the controller once.

ii. Only a fraction of the external traffic $\lambda_1$ and **not** a fraction of the net input traffic $\Gamma_1$ will go the controller. (Two lines directly out of the data node in Fig. 1(b) as opposed to one line in Fig. 1(a) are used to represent this phenomenon).

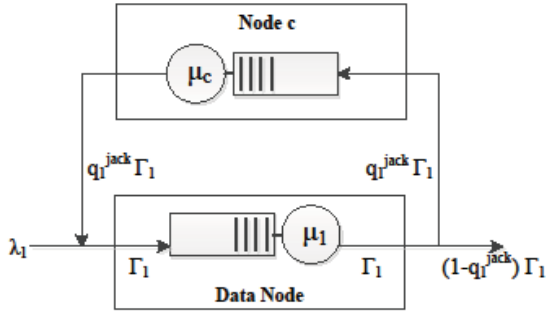
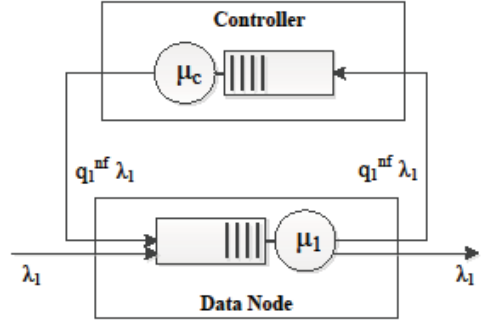

Figure 1(a): Jackson Model          Figure 1(b): Simple OpenFlow Model [1]

In an OpenFlow network, let $q_1^{nf}$ be the probability that the packet goes to the controller in case there is no flow entry in flow-table of the node, then in order to use the Jackson network to represent the OF-based SDN we have to adjust $q_1^{jack}$ by demanding that the input rates to the nodes in both the models are the same. Hence

$$\Gamma_1 = \lambda_1 + q_1^{nf} \lambda_1 \qquad (2)$$

and

$$q_1^{jack} \Gamma_1 = q_1^{nf} \lambda_1 \qquad (3)$$

As a result $q_1^{jack}$ can be solved as

$$q_1^{jack} = \frac{q_1^{nf}}{1 + q_1^{nf}} \qquad (4)$$

Fig. 2 highlights the need of having a *modified Jackson model* to represent OF-based SDN networks where we represent mean time spent by a packet in the network (node + controller) as a function of load on the controller.

The curve *simulation* is obtained from simulating the OpenFlow behavior taking into account the aforementioned two salient features. Further, in the simulation, we assume Poisson arrivals at the input and exponentially distributed service times for the nodes.

The curve denoted by **Jackson Model** is obtained by using $q_1^{jack}$ as such without modification i.e. $q_1^{jack} = q_1^{nf}$ while the curve **Modified Jackson Model** is based on $q_1^{jack}$ in (4).

It can be seen that as the percentage of traffic going to the controller dictated by $q_1^{nf}$ increases, the modification to the probability $q_1^{jack}$ in (4) becomes all the more important.

## 2.1. Limitations

The work in this paper makes the following assumptions:

- The overall traffic arrival process at the switch and the controller is Poison. Further exponentially distributed service times are used for the switch and the controller. This allows us to use the Jackson network results based on M/M/1 queues.

- Secondly we assume a single queue at the switch instead of a separate queue per line card.

- TCP traffic is used for which only the first packet of the unknown flow is sent to the controller.
- Infinite buffer is assumed at the switch as typically it is quite large.

It needs to be emphasized that the main goal of this work is to develop an analytical model for OF-based SDN networks. The assumptions will be relaxed in the subsequent work.

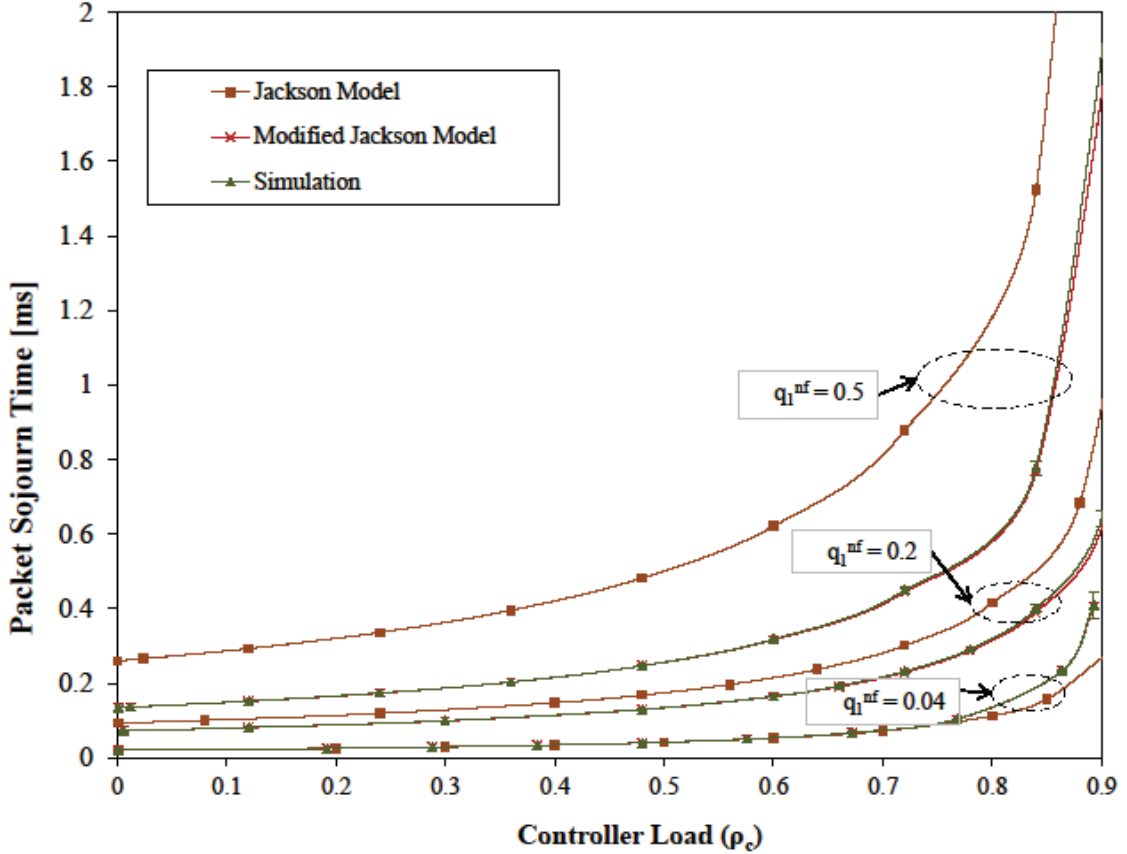

Figure 2: Jackson model cannot be used as such

## 3. PERFORMANCE MEASURES

One of the advantages of our proposed model is that we can leverage the well-established results for performance analysis of Jackson networks for analysing OpenFlow-based SDNs. In this section, we will use the proposed model to find two elementary performance measures, the average packet sojourn time, and the distribution of time spent by a packet in the network.

### 3.1. Average Packet Sojourn Time

The average packet sojourn time, $E[W^{jack}]$, is defined as the time spent by a packet in the network from the moment it enters the network at its source node, until it leaves through the destination. $E[W^{jack}]$ for the network in Fig. 1 is given as [11]

$$E[W^{jack}] = \frac{1}{\lambda_1} \left( \frac{\rho_1}{1 - \rho_1} + \frac{\rho_c}{1 - \rho_c} \right) \qquad (5)$$

where $\rho_1 = \Gamma_1/\mu_1$ and $\rho_c = \Gamma_c/\mu_c$ denote the load on the node *1* and the controller *c*, respectively. Further, in order to have a stable system, it is assumed that all the loads are less than unity that is $\rho_1 < 1$ and $\rho_c < 1$.

Alternatively, we can use the delay formula derived explicitly for the OpenFlow model, depicted in Fig. 1(b), as highlighted in [1].

To this end it needs to be highlighted that a packet arriving at the switch in the data plane of the OpenFlow network is confronted with two conditions. If there is already a flow entry installed in the switch then the packet is forwarded as such after spending time $T_1$ otherwise it goes to the controller, spends time $T_c$ then returns back to the same switch for packet matching where it spends time $T_1^{(2)}$ and is forwarded on the output interface.

So the absolute value of packet sojourn time $W^{of}$ in an OF-based SDN network where node $1$ interacts with the SDN controller $c$ as shown in Fig. 1(b) is given as

$$W^{of} = \begin{cases} T_1 & \text{with probability } 1 - q_1^{nf} \\ T_1 + T_c + T_1^{(2)} & \text{with probability } q_1^{nf} \end{cases} \quad (6)$$

where $T_1$ and $T_c$ are sojourn times in node $1$ and node $c$ respectively, while $T_1^{(2)}$ is the sojourn time when a packet enters node $1$ the second time after visiting the controller.

Eventually, the mean of $W^{of}$ is given as

$$E[W^{of}] = (1 - q_1^{nf}) E[T_1] + q_1^{nf} \left( E[T_1] + E[T_c] + E[T_1^{(2)}] \right)$$
$$= (1 + q_1^{nf}) \frac{1}{\mu_1 - \Gamma_1} + q_1^{nf} \frac{1}{\mu_c - \Gamma_c} \quad (7)$$

We draft a short proof in Lemma 1 to show that the mean packet sojourn time calculated by the two methods is indeed the same.

**Lemma 1:** For the single node case the packet sojourn time calculated in (5) using the standard Jackson assumption is the same as explicitly calculated using (7).

**Proof:** By rearranging (7) in terms of traffic loads we have:

$$E[W^{of}] = \frac{1 + q_1^{nf}}{\Gamma_1} \left( \frac{\rho_1}{1 - \rho_1} \right) + \frac{q_1^{nf}}{\Gamma_c} \left( \frac{\rho_c}{1 - \rho_c} \right) \quad (8)$$

Using $\Gamma_1 = (1 + q_1^{nf}) \lambda_1$ from (2) and $\Gamma_c = q_1^{nf} \lambda_1$ from (3) we obtain $E[W^{jack}]$, in (5) which proves the Lemma.

### 3.2. Distribution of Time Spent by the Packet

In this section we take a step forward by presenting the probability density function (PDF) and the cumulative density function (CDF) of the time spent by a packet in the node.

**Lemma 2:** The PDF $w_1^c(t)$ and the CDF $\widetilde{W}_1^c(t)$ of the time spent by a packet in the node $1$ are given respectively as

$$w_1^c(t) = b_1^{(1)} a_1 e^{-a_1 t} + b_1^{(2)} a_1 (a_1 t) e^{-a_1 t} + d_1 a_c e^{-a_c t} \quad (9)$$

$$\widetilde{W}_1^c(t) = P(W^{of} > t) = \left( b_1^{(1)} + b_1^{(2)} \right) e^{-a_1 t} + b_1^{(2)} (a_1 t) e^{-a_1 t} + d_1 e^{-a_c t} \quad (10)$$

where

$a_1 = \mu_1 - \Gamma_1, \ a_c = \mu_c - \Gamma_c$

while

$$b_1^{(1)} = 1 - q_1^{nf} - q_1^{nf} \frac{a_1 a_c}{(a_c - a_1)^2}, \quad b_1^{(2)} = q_1^{nf} \frac{a_c}{a_c - a_1}, \quad d_1 = q_1^{nf} \frac{a_1^2}{(a_c - a_1)^2}$$

**Proof:** If we assume that the sojourn times $T_1$ and $T_1^{(2)}$ are independent then the Laplace transform $W_1^c(s) = E[e^{-sW_1^{of}}]$ may be written as

$$W_1^c(s) = (1 - q_1^{nf}) \frac{a_1}{a_1 + s} + q_1^{nf} \left(\frac{a_1}{a_1 + s}\right)^2 \left(\frac{a_c}{a_c + s}\right) \tag{11}$$

which can further be written as

$$W_1^c(s) = b_1^{(1)} \frac{a_1}{a_1 + s} + b_1^{(2)} \left(\frac{a_1}{a_1 + s}\right)^2 + d_1 \left(\frac{a_c}{a_c + s}\right) \tag{12}$$

Inverting the Laplace transform proves the Lemma.

## 4. NUMERICAL RESULTS

In order to verify our proposed model we developed a discrete event simulation model to mimic the queuing behavior in an OF-based SDN. We assume that at the arrival to the node *1*, packets are queued in the data node before being processed. The processing time of data node $1/\mu_1$ is considered to be exponentially distributed with a mean value of 9.8μs. The value of 9.8 is the average processing time taken by Pronto 3290 switch for forwarding packets of size 1500 bytes [1]. We here assume that TCP uses maximum transmission unit (MTU) of 1500 bytes.

At the controller, the number of responses per second are taken to be 4175 as reported in [1] by using the Cbench tool [10]. Hence this is parameterized as $1/\mu_c = 240$μs in the model.

To enhance confidence in the simulation result, five replications for each value were run and the normally distributed 95% confidence interval is incorporated in the plots.

We first highlight in Fig. 3 that our proposed model provides a fix to the results reported by [1]. To this end we first plot the ***Simulation*** curve from [1] as a reference. On top of it we plot the analytical curves; ***Analytical***[1]} and ***Modified Jackson Model***, obtained using the model in [1] and our proposed Jackson model, respectively.

It can be seen that the model proposed by [1] performs very well for small loads ($q_1^{nf}$=0.2). However in the cases when there is a large amount of query traffic coming to the controller due to unknown flows the model in [1] falls short. In such cases the proposed modification to the Jackson model is quite accurate as seen for the extreme case of $q_1^{nf}$=1.0.

In Fig. 4, the effect of $q_1^{nf}$ on network throughput is studied where the network throughput is defined as the amount of traffic $\lambda_1$ which can be injected into the OF-based SDN for a given delay guarantee. In this case the delay guarantee is the average packet sojourn time. This plot also highlights how the proposed model can be used to dimension the network if packet sojourn time is considered as the design parameter. A striking feature of this plot is that the network throughput saturates after reaching a certain value of packet sojourn time. Subsequently, it can be inferred that even if the network is over-loaded after crossing a certain traffic threshold, the result will be just increased packet sojourn time without further enhancing the network throughput. Similarly, it is also observed that the critical value for packet sojourn time remains almost the same for all the values of $q_1^{nf}$ but the resulting network throughput for each of them is quite different.

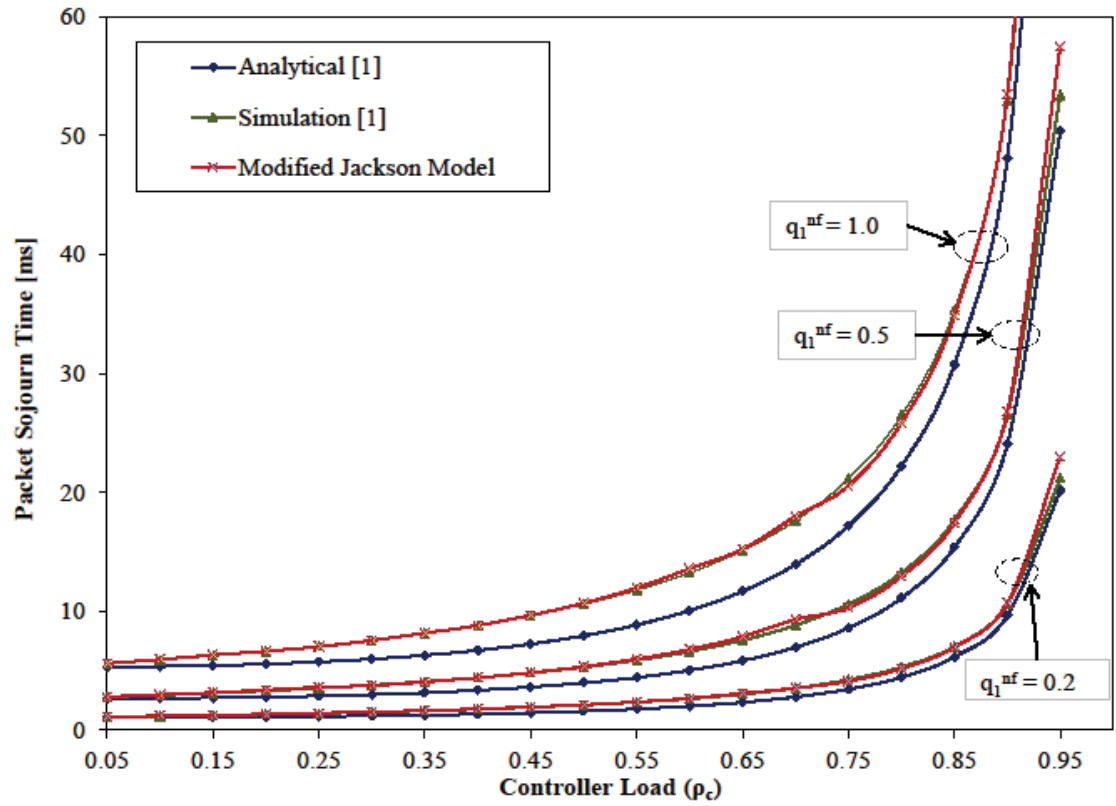

Figure 3: Comparing Modified Jackson Model to Results from [1]

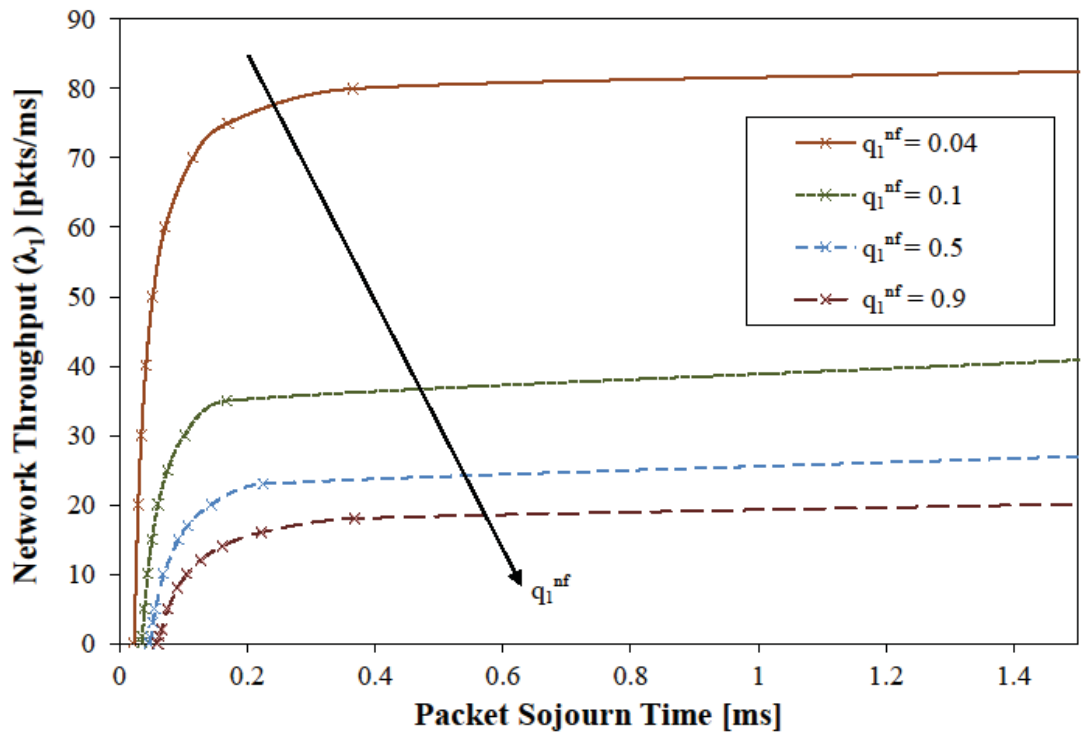

Figure 4: Dimensioning Network Throughput

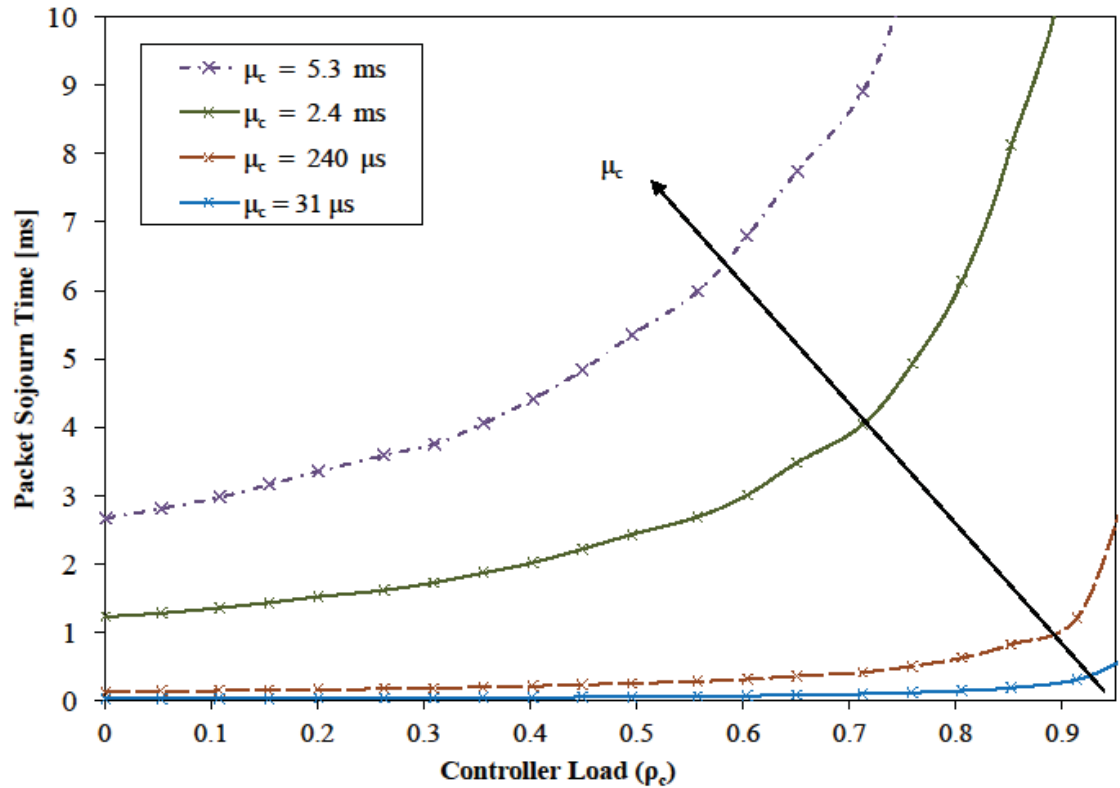

Figure 5: Effect of the controller service rate $\mu_c$

In Fig. 5 a fundamental performance plot is shown in which packet sojourn time is plotted against the controller load $\rho_c$ for differing values of the controller service time $\mu_c$ with $q_1^{nf}$ constant at 0.5. Although the plot is mainly for evaluating performance, it can also play a role in designing a network with controller of known average service time and giving guarantees on packet sojourn time by keeping the controller load at a certain level.

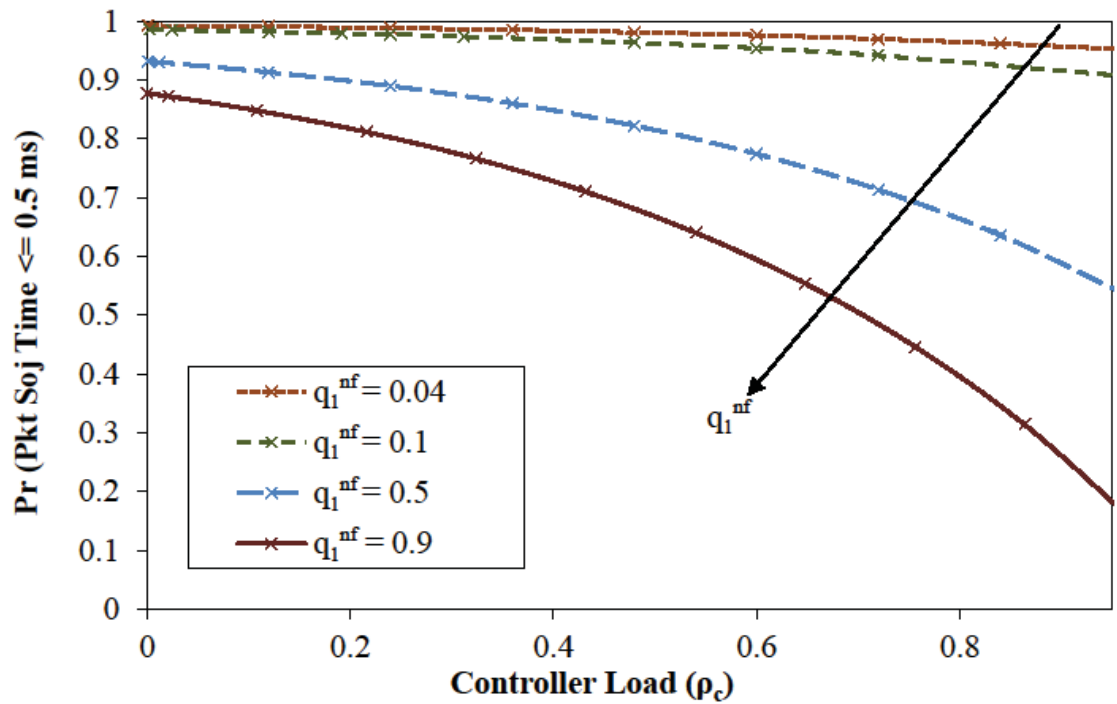

Figure 6: CCDF of packet sojourn time as a function of load on controller

In Fig. 6, the plot shows the probability that packet sojourn time ≤ 0.5 ms for varying values of controller load $\rho_c$ and for different $q_1^{nf}$ values. The plot can be used to determine the maximum load that the controller should reach before its performance is compromised. The plot in Fig. 6 is pilot and similar plots for different values of packet sojourn time can be obtained depending upon the requirements.

It needs to be emphasized that blocking probability $p_b$ was zero for the setup which we had for the simulation and infinitesimal small for the analytical model.

## 5. THE MULTI-NODE CASE

In a real life SDN deployment, an SDN controller is responsible for more than one node in the data plane. In this section we highlight how the proposed model can be used to model this scenario. To this end we take a toy example in which we only have two nodes in the data plane as shown in Fig. 8. We define $q_2^{jack}$ and $q_2^{nf}$ for node *2* similar to $q_1^{jack}$ and $q_1^{nf}$ defined earlier for node *1*.

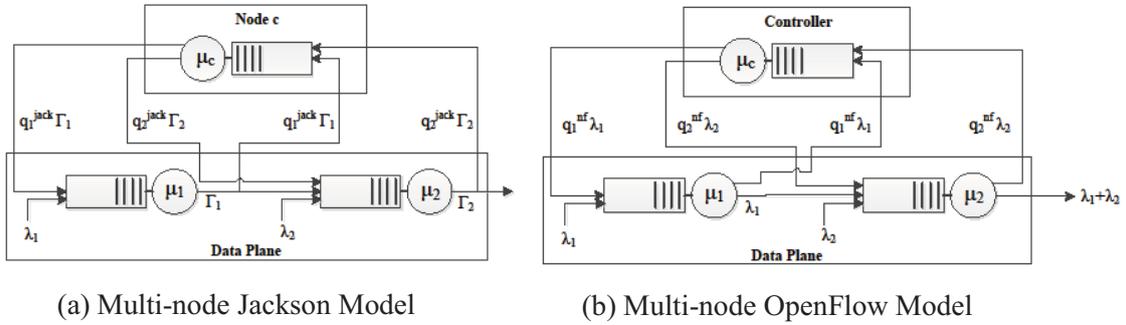

(a) Multi-node Jackson Model  (b) Multi-node OpenFlow Model

Figure 7: Modeling more than one node in the data plane

In order to leverage the Jackson model in Fig. 8(a) for modeling the OF-based SDNs in Fig. 8(b), the probabilities $q_2^{jack}$ and $q_1^{jack}$ need to be adjusted. This is accomplished by forcing the rates at all the nodes in both the models to be the same as

For node *1*:  $\lambda_1(1+q_1^{nf}) = \Gamma_1, \quad q_1^{nf}\lambda_1 = q_1^{jack}\Gamma_1$  (17)

For node *2*:  $\lambda_1+\lambda_2(1+q_2^{nf}) = \Gamma_2, \quad q_2^{nf}\lambda_2 = q_2^{jack}\Gamma_2$  (18)

Solving (17) we get $q_1^{jack}$ same as (4) while by solving (18) we have $q_1^{jack}$ as

$$q_2^{jack} = \frac{q_2^{nf}\lambda_2}{\lambda_1+\lambda_2(1+q_2^{nf})} \quad (19)$$

We can then use the modified $q_1^{jack}$ and $q_2^{jack}$ to derive the appropriate performance metrics such as packet sojourn time using existing queuing theory results [11] similar to the single node case.

## 6. CONCLUSION

In this work we have proposed an analytical model for an OpenFlow enabled SDN based on Jackson network. We have shown that the model is accurate even for the case when the probability of new flows is quite large. The applicability of the model is determined by establishing two performance measures, the average packet sojourn time and the distribution of time spent by a packet in the network, by using the proposed model. Secondly we showed by a toy example that the model can be readily extended to more than one switch in the data plane. Conclusively it is noted, and can be safely stated, that the model proposed in this paper caters for realistic OpenFlow-based SDNs and this argument has readily been validated in this paper.

Furthermore, the effects of key parameters in an SDN network are studied which include the time required by the controller to process a request, amount of traffic going to the controller, average time spent by a packet in a network and the network throughput.

There is more than one direction that the work presented in this paper can be taken forth. First of all, the work presented and validated for a single node can be extended to larger and more realistic topological scenarios, such as fat-tree topology. Secondly, the model in this work is based on Markovian arrival and service processes which can be generalized and more realistic distributions or traces can be used in modeling. This can be supplemented with simulations for validation and verification of the model. Also, a test-bed study for verifying our model can be performed which will enhance the confidence in the proposed model.